\documentclass[]{revtex4}
\usepackage{float}
\usepackage{amsmath}
\usepackage{amssymb}
\usepackage{amsfonts}
\usepackage{euscript}
\usepackage{enumerate}
\usepackage{hhline}
\usepackage[para,online,flushleft]{threeparttable}
\usepackage{pslatex}
\usepackage{tabularx}
\usepackage[usenames,dvipsnames]{xcolor}
\usepackage{graphicx}
\usepackage{dcolumn}
\usepackage{bm}
\usepackage[sort&compress]{natbib}

\usepackage{lipsum}

\newcommand{\epsrr}{\varepsilon\left(\boldsymbol{\rho}, \boldsymbol{r}\right)}
\newcommand{\SiN}{\text{Si}_\text{3}\text{N}_\text{4}}
\newcommand{\mum}{ \,\mu\text{m}}

\begin{document}

\title{Multi-objective Inverse Design of Solid-state Quantum Emitter Single-photon Sources}
\author{Emerson G. Melo}
\email{emerdemelo@usp.br}
\affiliation{Materials Engineering Department, Lorena School of Engineering, University of S\~{a}o Paulo, Lorena, SP 12602-810, Brazil}
\affiliation{National Institute of Standards and Technology, Gaithersburg, MD, 20899 USA}
\author{William Eshbaugh}
\affiliation{Department of Physics and Astronomy, West Virginia University, Morgantown, WV, 26506 USA}
\author{Edward B. Flagg}
\affiliation{Department of Physics and Astronomy, West Virginia University, Morgantown, WV, 26506 USA}
\author{Marcelo Davanco}
\email{marcelo.davanco@nist.gov}
\affiliation{National Institute of Standards and Technology, Gaithersburg, MD, 20899 USA}



\begin{abstract}
Single solid-state quantum emitters offer considerable potential for the implementation of sources of indistinguishable single-photons, which are central to many photonic quantum information systems. Nanophotonic geometry optimization with multiple performance metrics is imperative to convert a bare quantum emitter into a single-photon source that approaches the necessary levels of purity, indistinguishability, and brightness for quantum photonics. We present an inverse design methodology that simultaneously targets two important figures-of-merit for high-performance quantum light sources: the Purcell radiative rate enhancement and the coupling efficiency into a desired light collection channel. We explicitly address geometry-dependent power emission, a critical but often  overlooked aspect of gradient-based optimization of quantum emitter single-photon sources. We illustrate the efficacy of our method through the design of a single-photon source based on a quantum emitter in a GaAs nanophotonic structure that provides a Purcell factor $F_p=21$ with a 94~\% waveguide coupling efficiency, while respecting a geometric constraint to minimize emitter decoherence by etched sidewalls. Our results indicate that multi-objective inverse design can yield competitive performance with more favorable trade-offs than conventional approaches based on pre-established waveguide or cavity geometries. 
\end{abstract}

\maketitle

\section{Introduction}

Sources of indistinguishable single photons are central to a number of photonic quantum information systems currently being developed for quantum computing, simulation, communication, sensing, and metrology~\cite{Eisaman2011,Senellart.Pascale.2017}. Single solid-state quantum emitters~\cite{Aharonovich.Igor.2016} offer considerable potential for the implementation of such sources, for two primary reasons. First, single photons can be produced on demand, at rates that can readily reach the gigahertz range, being fundamentally limited only by the characteristic cycling times between the emitter's excited and ground states. Second, solid-state quantum emitters can be embedded into nanophotonic structures that can provide highly efficient and selective photon funneling into desired spatial modes~\cite{ref:Davanco_BE, mouradian_scalable_2015,davanco_heterogeneous_2017, katsumi_quantum-dot_2019}, or allow control of the emitter's radiative rate via the Purcell effect~\cite{Lodahl2015}. By combining these two capabilities one can convert a quantum emitter with favorable intrinsic properties into a single-photon source that approaches ideal levels of single-photon purity, indistinguishability, and brightness~\cite{senellart_high-performance_2017}. 

High Purcell radiative rate enhancement factors ($F_p$) are of crucial importance for single-photon sources based on any type of quantum emitter, as the effect can be leveraged to achieve higher single-photon generation rates\cite{Birowosuto2012}, improved photon indistinguishability~\cite{Liu2018, somaschi_near-optimal_2016}, and higher $\beta$-factors~\cite{Kaer.P.2013}. The $\beta$-factor accounts for the probability that a photon produced by an optical transition couples into a specific optical spatial mode of the quantum emitter's host geometry, such as a cavity resonance or a bound waveguide mode. Frequently, photons coupled to such spatial modes must be extracted into a desired collection channel, such as an optical fiber~\cite{ref:Davanco2,Davanco2011,Schroeder2011a,daveau_efficient_2017} or on-chip waveguide mode~\cite{mouradian_scalable_2015,davanco_heterogeneous_2017,katsumi_transfer-printed_2018}. If extraction is accomplished with an efficiency $\kappa$, the total single-photon coupling efficiency into the desired collection channel is then $\eta=\beta\kappa$ \cite{katsumi_transfer-printed_2018}. As a consequence, although some level of Purcell radiative rate enhancement may be expected when employing only $\eta$ as a figure-of-merit (FOM), or, conversely, an increase in overall coupling efficiency may be expected when only $F_{p}$ is targeted, there is no guarantee in either case that both parameters will be maximized simultaneously, as is required for sources that need both high efficiency and indistinguishability~\cite{Aharonovich.I.2016}. To achieve sources with favorable trade-offs between $F_p$ and $\eta$, therefore, a nanophotonic geometry optimization strategy that addresses multiple performance metrics is imperative. 

Many optimization approaches have recently been developed for nanophotonic design~\cite{Mao_algorithms_2021}. In particular, shape and topology optimization techniques based on the adjoint method have been applied to the inverse design of nanophotonic devices~\cite{C.M.Lalau.2013_invdes,Sean.Molesky.2018}. Such a strategy has become very attractive because it promises a simplified, goal-oriented approach to device design and optimization. In this method, a spatially varying refractive index distribution is allowed to evolve to optimize a pre-defined device response. The response is modeled as a user-defined cost function that includes both desirable optical characteristics and fabrication constraints. The adjoint method enables efficient calculation of the relevant gradient of the cost function by solving only two related constraint equations no matter how many design parameters there are. This inverse-design strategy has been applied successfully to produce a variety of integrated photonic elements such as non-adiabatic waveguide tapers~\cite{Michaels.Andrew.2018_continuous_design}, beam splitters~\cite{Michaels.Andrew.2020}, grating couplers~\cite{Michaels.Andrew.2018}, filters\cite{Julian.L.Pita.2019,Su.Logan.2018}, photonic crystals~\cite{Vercruysse.Dries.2020_invd_phc}, metasurfaces~\cite{Chawin.Sitawarin.2018}, laser-driven particle accelerators~\cite{Neil.V.Sapra.2020_particle_accelerator}, and micro- and nanocavities~\cite{X.Liang.2013_invdes_nanocavity, Jesse.Lu.2010_invd_convex_opt,Jesse.Lu.2011_invd_3d_resonator, Ahn.G.H.2022_invd_microresonator}.

\begin{figure*}[ht]
	\centering
	\includegraphics[scale=1.2]{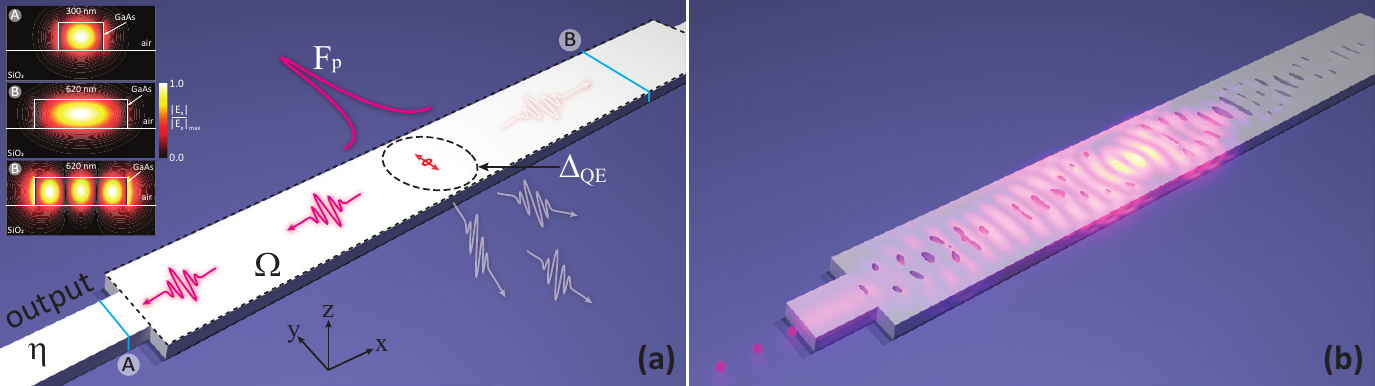}
	\caption{Single-photon source geometric optimization problem. \textbf{(a)} Single-photons emitted by a quantum emitter, modeled as an electric dipole source inside of a domain $\Omega$, must be funneled into the fundamental optical mode of an output waveguide, whereas coupling into other channels must be suppressed. The proposed inverse-design methodology simultaneously optimizes the Purcell radiative rate enhancement factor ($F_{p}$) and the overall coupling efficiency ($\eta$) for the single emitter by varying the spatial permittivity distribution within the design region $\Omega$. While this problem formulation is general, the specific example of a single InAs quantum dot in a GaAs host is used in this work for illustration. The domain $\Omega$ in this case consists of a GaAs waveguide of 620~nm width and 190~nm height on a SiO$_2$ substrate, which supports four transverse-electric (TE) modes, two of which are shown in the figure insets. The dipole is placed at the center of $\Omega$ in the y- and z-directions. The output waveguide has a width of 300~nm, thereby supporting only one fundamental TE optical mode. A constraint of the optimization is that permittivity variations are disallowed within a circular region of radius $\Delta_{QE} = 300$ nm around the emitter, as indicated. This will reduce the negative effects of etched sidewalls on the quantum dot. \textbf{(b)} Artistic rendering of the single-photon source design obtained with the proposed optimization methodology for which the values of $F_{p}$ and $\eta$ are respectively 21 and 0.94.}
	\label{fig1:device}
\end{figure*}

In this theoretical work, we develop an inverse design methodology that simultaneously targets the two main FOMs for high-performance quantum light sources---the Purcell factor $F_{p}$ and the overall quantum emitter coupling efficiency $\eta$ into a desired collection channel---stated as a multi-objective optimization problem. We apply our method to the design of a waveguide-coupled single-photon source based on a single InAs/GaAs quantum dot emitter, with a geometrical constraint that minimizes adverse effects on the latter cause by sidewall proximity~\cite{Liu.Jin.2018}. We show that efficiencies, $\eta$, comparable or superior to those obtained with conventional design approaches based on pre-established waveguide or cavity geometries~\cite{katsumi_transfer-printed_2018} can be achieved with simultaneously high $F_p$ values, in spite of the geometrical constraint. Our results show that inverse design procedures that target either only $\eta$ or only $F_{p}$~\cite{Raymond.A.W.2021_nv_sps,Srivatsa.C.2020_nv_sps,Ian.M.Hammond.2021_invd_meep,Yesilyurt.O.2021_sps_coupler} may lead to sub-optimal $F_{p}$ or $\eta$, respectively, whereas multi-objective optimization produces highly favorable trade-offs between the two parameters regarding single-photon source performance. To obtain these results, we employ a formulation of the adjoint simulation source that, critically, accounts for the variation of the total power emitted by the quantum emitter due to the Purcell effect~\cite{Lodahl2015,X.Liang.2013_invdes_nanocavity}. Such consideration is not only necessary for the precise evaluation of optimization gradients to allow efficient convergence, but is also shown to produce subtle, advantageous trade-offs that have not been observed in prior investigations~\cite{Raymond.A.W.2021_nv_sps,Srivatsa.C.2020_nv_sps,Ian.M.Hammond.2021_invd_meep}. 

While we show results for a specific type of quantum emitter and material host, our methodology is general and applicable to any nanophotonic system in which an efficient optical access interface to a single quantum emitter must be created and high indistinguishability is a requirement. 

\section{Methodology}
\label{section:methodology}
Various implementations of adjoint-based inverse design frameworks have been built so far~\cite{C.M.Lalau.2013_invdes,Michaels.Andrew.2018_continuous_design,Julian.L.Pita.2019,Su.Logan.2020_spins,H.Tyler.W.2018_invdes_nonlinear,Minkov.M.2020_ceviche}. Regardless of the numerical solver and  parametrization strategy (e.g. shape and topology) chosen to control the refractive index distribution, some common aspects of the optimization framework must be properly addressed to achieve an efficient final design when employing such a tool. We next describe how the setup of objective functions, adjoint sources, and geometrical constraints may be addressed for the optimization of single-photon sources based on single quantum emitters.

The problem studied in this work is schematically depicted in Fig.~\ref{fig1:device}(a), where a quantum emitter is modeled as a y-oriented electric dipole source embedded in a domain $\Omega$, which we choose to be a multimode optical waveguide. The multimode waveguide is connected to an output waveguide supporting a single mode, which is the desired photon collection channel. By employing inverse design, we want to transform the design domain $\Omega$ such that (i) highly directional single-photon emission is obtained through the fundamental mode of the output waveguide, i.e., the coupling into other guided or radiation modes is suppressed; and (ii), a maximum Purcell factor is achieved.

The photonic geometry optimization is first cast as a multi-objective minimization problem stated as~\cite{Su.Logan.2018}
\begin{equation}
	\label{eq:opt_problem}
	\begin{aligned}
		& \underset{\boldsymbol{\rho}}{\text{minimize}}
		& & F(\boldsymbol{x}(\boldsymbol{\varepsilon}(\boldsymbol{\rho}))) = f_{1} + f_{2} =  \left(\frac{1}{F_{p}^{\theta}} - \frac{1}{F_{p}\left(\boldsymbol{x}\right)} \right)^{2} + \left(\eta_{m}^{\theta} - \eta_{m}\left(\boldsymbol{x}\right)\right)^2 \\
		& \text{subject to}
		& & \nabla \times \frac{1}{\mu_{0}} \nabla \times \boldsymbol{E}\left(\omega, \boldsymbol{r}\right) - \omega^{2}\varepsilon\left(\boldsymbol{\rho}, \boldsymbol{r}\right)\boldsymbol{E}\left(\omega, \boldsymbol{r}\right) = -i\omega \boldsymbol{J}\left(\omega, \boldsymbol{r}\right)
	\end{aligned}
\end{equation}
with $\boldsymbol{x}=[\boldsymbol{E}\left(\omega, \boldsymbol{r}\right),  \boldsymbol{H}\left(\omega, \boldsymbol{r}\right)]^{T}$. Here, $\boldsymbol{E}$ and $\boldsymbol{H}$ are the electric and magnetic fields  produced by a current dipole source $\boldsymbol{J}$ oscillating at the angular frequency $\omega$ in a medium with permittivity $\varepsilon$ and free-space permeability $\mu_{0}$. The permittivity $\varepsilon\left(\boldsymbol{\rho}, \boldsymbol{r}\right)$ has spatial dependence described by a set of design parameters $\boldsymbol{\rho} \in [0, 1]$. Of critical importance is the implicit dependence of the fields, $\mathbf{x}$, on the permittivity distribution---and thus the design parameters---via the constraint relation in the second line of Eq.~\ref{eq:opt_problem}. The cost function $F(\boldsymbol{x})$ in Eq.~\ref{eq:opt_problem} captures a desired optical response, calculated from $\boldsymbol{E}$ and $\boldsymbol{H}$, which is optimized via the design parameters $\boldsymbol{\rho}$. We define $F(\boldsymbol{x})$ as a sum of two objectives, $f_{1}$ and $f_{2}$, to be optimized simultaneously, respectively involving the Purcell factor $F_{p}$ and a modal coupling efficiency $\eta_{m}$. Here, the superscript $\theta$ indicates desired values of target quantities (ideally, $F_{p}^{\theta} \to \infty$ and $\eta^{\theta} = 1$), and $m = 0, 1 , 2 \dots, M$ refers to the output waveguide optical mode intended for single-photon extraction. In the case illustrated below, we use $m = 0$. It is important to note that the current dipole, $J\left(\omega, \boldsymbol{r}\right) = \delta\left(\boldsymbol{r}-\boldsymbol{r'}\right)\boldsymbol{e}_{j}$, located at $\mathbf{r}'$, radiates a power that is inherently dependent on the design parameters $\boldsymbol{\rho}$, since the source is placed inside the design domain region $\Omega$, as indicated in Fig.~\ref{fig1:device}(a). This design-dependent radiated power must be addressed to correctly calculate both the Purcell factor and the modal coupling efficiencies.

A critical need of gradient-based optimization engines is the accurate and efficient calculation of the derivatives of $F(\boldsymbol{x})$ with respect to the design variables~\cite{Michaels.Andrew.2018_continuous_design,H.Tyler.W.2018_invdes_nonlinear}. Following the adjoint method, such derivatives can be accurately calculated from electromagnetic field distributions obtained from two sequential simulations, a forward one and an adjoint one~\cite{Michaels.Andrew.2018_continuous_design}. In the formalism of ref.~\cite{Michaels.Andrew.2018_continuous_design},  the sensitivities $\partial F(\boldsymbol{x}) / \partial \rho_{n}$ can be expressed as 
\begin{equation}
	\label{eq:grad_F}
	\frac{\partial F(\boldsymbol{x})}{\partial \rho_{n}} = -2 Re\left\{\boldsymbol{y}^{T} \frac{\partial \boldsymbol{A}}{\partial \rho_{n}} \boldsymbol{x}\right\}.
\end{equation}
In Eq.~(\ref{eq:grad_F}), $\boldsymbol{x}$ and $\boldsymbol{y}$ are the forward and adjoint fields in the design region $\Omega$, respectively, and the matrix $\boldsymbol{A}$ results from a finite-differences discretization of Maxwell's equations with the spatially varying permittivity $\varepsilon\left(\boldsymbol{\rho}, \boldsymbol{r}\right)$. For the single-photon source problem, the forward field $\boldsymbol{x}$ is obtained by simulating an electric dipole point source radiating in a medium with an initial  $\epsrr$, which amounts to solving the matrix equation $\boldsymbol{A}\boldsymbol{x} = \boldsymbol{b}$, where the vector $\boldsymbol{b}$ represents the dipolar source. The adjoint fields $\boldsymbol{y}$ are  obtained by solving the equation $\boldsymbol{A}^{T}\boldsymbol{y} = \left(\partial F/\partial \boldsymbol{x}\right)^{T}$, which corresponds to simulating the fields produced by a current source defined by  the derivative of $F(\boldsymbol{x})$ with respect to the electric and magnetic field components of the forward field $\boldsymbol{x}$. We now show explicit expressions for such derivatives, considering the composite cost function $F(\boldsymbol{x})$ in Eq.~($\ref{eq:opt_problem}$).

The Purcell factor $F_p$ in Eq.~(\ref{eq:opt_problem}) may be obtained as the ratio $F_p={LDOS_{D}}/{LDOS_{B}}$ between the per-polarization local electromagnetic density of states (LDOS) in the nanophotonic device (D) and in bulk semiconductor (B). Following the definition in~\cite{X.Liang.2013_invdes_nanocavity}, the device or bulk LDOS can be calculated as
\label{eq:LDOS}
\begin{equation}
LDOS\left(\omega, \boldsymbol{r'}\right) = -\frac{6}{\pi}Re\left[\int\boldsymbol{J}^{\ast}\left(\omega, \boldsymbol{r}\right)\cdot\boldsymbol{E}\left(\omega, \boldsymbol{r}\right) d^{3}\boldsymbol{r}\right].    
\end{equation}
That stated, the derivative of $f_1$ in Eq. (\ref{eq:opt_problem}) with respect to the electric field components $E_{\phi,ijk}$ ($\phi = x, y, z$) evaluated at each position $(i,j,k)$ on a discretized three-dimensional spatial grid is
\begin{equation}
	\label{eq:df1_dE}
	\frac{\partial f_{1}}{\partial E_{\phi,ijk}} = -2\left(\frac{1}{F_{p}^{\theta}}-\frac{1}{F_{p}}\right)\left(\frac{\frac{3}{\pi}LDOS_{B}\delta_{V} J_{\phi,ijk}^{\ast}\left(\omega\right)}{\left\{\frac{6}{\pi}\delta_{V} Re\left[\boldsymbol{J}^{\ast}_{ijk}\left(\omega\right)\cdot\boldsymbol{E}_{ijk}\left(\omega\right) \right]\right\}^{2}}\right),
\end{equation}
where $\delta_{V}$ is the grid cell volume. The derivatives with respect to the magnetic field are all null in this case, so $\partial f_{1}/\partial H_{\phi,ijk} = 0$. $F_p$ may be equivalently calculated as the ratio $P_{D}/P_{B}$ of the dipole-radiated powers in the device and in bulk semiconductor respectively~\cite{vuckovic_finite-difference_1999}, which may provide a more intuitive understanding of the dipolar emission. Using the LDOS ratio for the $F_{p}$ definition, however, leads to simpler expressions in Eq.~\ref{eq:df1_dE}.

The waveguide mode coupling efficiency $\eta_{m}$ relates the total power $P_{0}$ emitted by the dipole source and the energy coupled into the output waveguide mode $m$. In the absence of reflected waves at the output collection waveguide, $\eta_{m}$ can be calculated as \cite{Michaels.Andrew.2018_continuous_design}
\begin{equation}
	\label{eq:eta}
	\eta_{m} = \frac{1}{4P_{0}P_{m}}\left|\iint_{A}d\boldsymbol{A}\cdot\boldsymbol{E}\left(\omega, \boldsymbol{r}\right)\times\boldsymbol{H}^{\ast}_{m}\left(\omega, \boldsymbol{r}\right)\right|^2,
\end{equation}  
where $\boldsymbol{E}$ is the electric field of the forward solution, $\boldsymbol{H}_{m}$ is the unnormalized magnetic field of mode $m$ of the output waveguide, and $P_{m}$ is the longitudinal component of the unnormalized mode's Poynting vector integrated over the cross-section, as highlighted in Fig.~\ref{fig1:device}(a). We emphasize that normalization by the source power $P_0$ in Eq.~\ref{eq:eta} is essential for the accurate calculation of the adjoint source, since $P_0$ is expected to change with the varying dielectric function $\epsrr$. Absence of such normalization will, in most cases, lead to incorrect sensitivities and, consequently, low optimization convergence. Equation ~\ref{eq:eta} may be expressed as a function of the modal complex expansion coefficient for the forward traveling wave, as $\eta_{m} = (|a_{m}|^{2}P_{m})/P_{0}$~\cite{O.Shapira.2005_mode_decomposition, Michaels.Andrew.2019_thesis}. In this case, the derivative of $f_{2}$ with respect to the field components at each grid position is 
\begin{equation}
	\label{eq:df2_dx}
	\frac{\partial f_{2}}{\partial x_{\phi,ijk}} = -2\left(\eta_{m}^{\theta} - \eta_{m}\right)\left(P_{m}\frac{\frac{da_{m}}{dx_{\phi,ijk}}a_{m}^{\ast}P_{0} - \frac{dP_{0}}{dx_{\phi,ijk}}|a_{m}|^{2}}{P_{0}^{2}}\right).
\end{equation} 
Here, the derivatives of $a_{m}$ and $P_{0}$ with respect to the field components can be obtained after employing Riemann sums to approximate integrals, writing out cross products, and using the Poynting vector and its complex conjugate to take the real part of the complex power, as detailed in \cite{Michaels.Andrew.2019_thesis}.

Besides a proper cost function and adjoint source evaluation, a judicious choice of initial geometry can favor the optimization process considerably~\cite{Michaels.Andrew.2020}, speeding up convergence, and potentially avoiding local minima. We next demonstrate the use of our methodology through an example that illustrates all such points. We optimize a single-photon source composed of a single InAs quantum dot embedded in a $190$ nm thick GaAs slab, a mature material system that has been extensively explored in quantum optics and quantum photonics~\cite{Dietrich2016}. We model the single InAs quantum dot as an electric current dipole emitting at a wavelength of 940~nm. Our choice of initial geometry, shown in Fig.~\ref{fig1:device}(a), takes into account two important considerations. First, the central portion of the geometry has a minimum width of 620~nm, so that the dipole is $>300$ nm from any etched surface. Such a constraint is desirable for minimizing quantum dot linewidth broadening caused by proximity to etched sidewalls, as observed in Ref.~\cite{Liu.Jin.2018}. To prevent the formation of sidewalls near the quantum emitter during the optimization process, the design parameters $\rho$ were constrained to a constant value of 1 inside a circle with a radius $\Delta_{QE}=300$~nm centered at the dipole (indicated by the dashed line in Fig.~\ref{fig1:device}(a)). The second consideration was that the output waveguide supports only a single transverse-electric (TE) mode, to which the coupling will be maximized. A width of 300~nm ensures that only a single TE mode is supported, shown in Fig.~\ref{fig1:device}(a). The result of the case illustrated here can be seen in Fig.~\ref{fig1:device}(b), whereas detailed analysis and discussion are provided in the next section.

We have implemented the Purcell factor and output waveguide coupling efficiency figures-of-merit within the framework of a pre-existing open-source nanophotonic inverse-design software package~\cite{Su.Logan.2020_spins}. An implementation for graphics processing units (GPUs) of the finite-difference frequency-domain (FDFD) method ~\cite{Su.Logan.2020_spins} was used to obtain the forward and adjoint electromagnetic fields. The material permittivity inside the design area was transformed following a density-based approach to topology optimization \cite{Jensen.J.S.2011_topology_nanophotonics}, i.e., the permittivity at each finite-difference grid cell was assigned to a design parameter as $\varepsilon(\tilde{\rho}) = \varepsilon_{min} + \tilde{\rho}(\varepsilon_{max} - \varepsilon_{min}), \tilde{\rho} \in [0, 1]$. Here, $\varepsilon_{min}$ and $\varepsilon_{max}$ are the minimum and maximum values of the permittivity, corresponding respectively to those of the air and the bulk semiconductor, and $\tilde{\rho}$ are filtered design parameters obtained after applying a tangent hyperbolic projection filter over $\rho$ to enforce binarization \cite{Wang.F.2011_hyperbolic_projection_filter}. The smoothness of this filter is controlled by a projection parameter that is increased every 5 iterations, leading to steeper transitions from GaAs to vacuum in the design region.

\section{Results and Discussions}
\begin{figure*}[ht]
	\centering
	\includegraphics[scale=1.1]{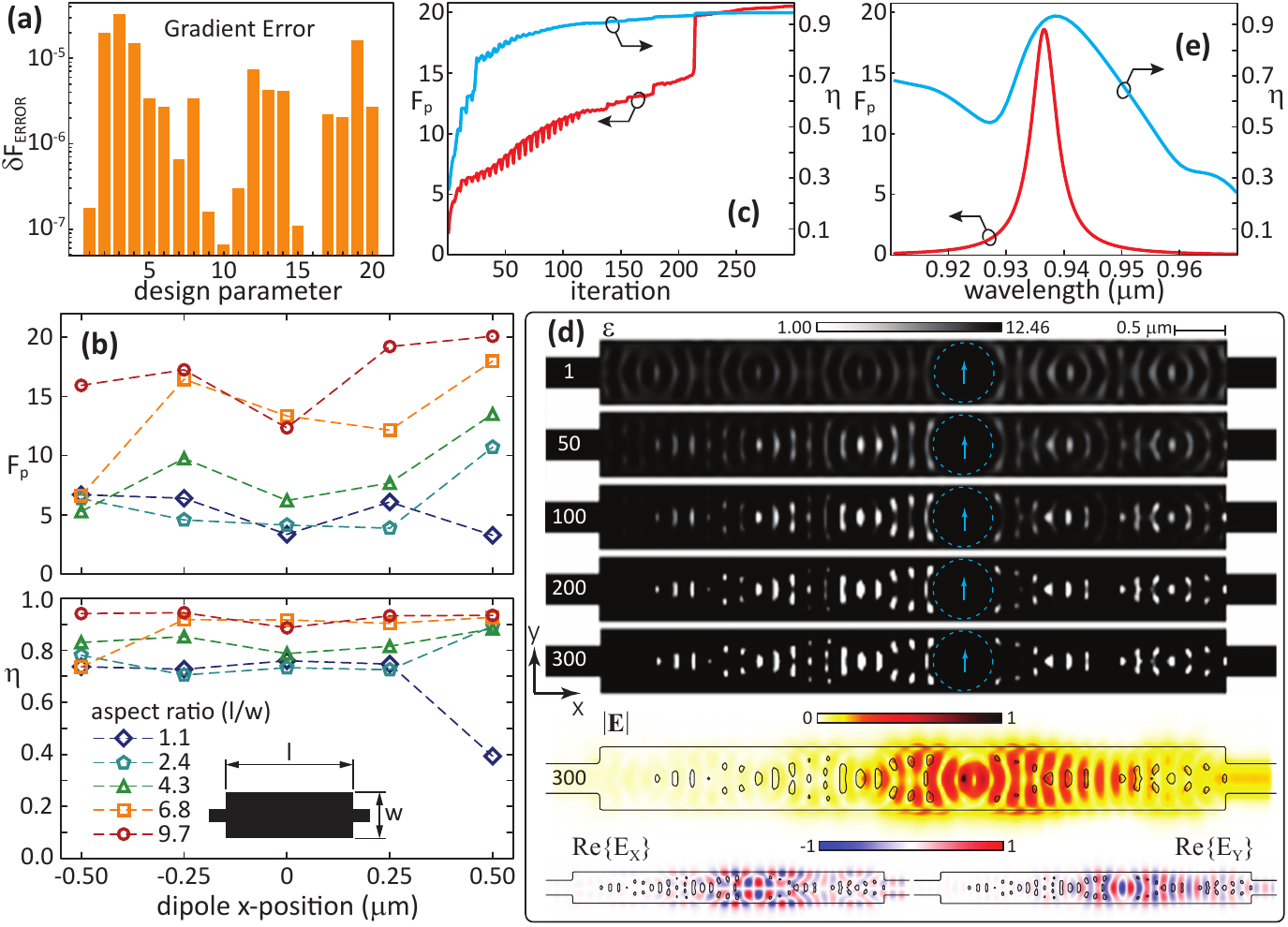}
	\caption{Optimization results. \textbf{(a)} Gradient error calculated for 20 randomly selected design parameters. \textbf{(b)} Final values of Purcell factor ($F_{p}$) and overall coupling efficiency ($\eta$) for optimizations considering devices with aspect ratios varying from 1.1 to 9.7 (as indicated in the bottom inset) and electric dipoles located at x-positions ranging from -0.5 $\mum$ to 0.5 $\mum$. Dashed lines are only guides for the eye; \textbf{(c-e)} Details concerning the optimization of the device with 9.7 aspect ratio (6 $\mum$ $\times$ 0.62 $\mum$) and the dipole source placed at the position $\boldsymbol{r'} = (0.5, 0, 0)\:\mum$, for which the values of $F_{p}$ and $\eta$ are respectively 21 and 0.94. \textbf{(c)} Values of $F_{p}$ and $\eta$ at each iteration; \textbf{(d)} dielectric distribution ($\boldsymbol{\varepsilon}$) at iterations 1, 50, 100, 200, and 300, the absolute value of electric field ($|E|$), and the real part of the electric field components in the x ($\operatorname{Re}\{E_{X}\}$) and y ($\operatorname{Re}\{E_{Y}\}$) directions. The region within the blue dashed circle of radius $\Delta_{QE}=300$ nm remains unchanged throughout the iterations. Blue arrows indicate the orientation of the electric dipole source. The electric field data were taken at the last iteration and are normalized by their maximum values. \textbf{(e)} Spectral curves of $F_{p}$ and $\eta$ as a function of wavelength, obtained by the finite-difference time-domain (FDTD) method.}
	\label{fig2:fp_eta_results}
\end{figure*}

As a first test of our method, an assessment of the gradient calculation accuracy was performed, where the $\delta F/\delta\rho_{n}$ value of $20$ $\rho_{n}$ design parameters, randomly chosen, was calculated through the adjoint ($\delta F_{ADJ}$) and finite-difference expressions ($\delta F_{FDM}$). The gradient error of the sensitivities obtained by the adjoint method with respect to the finite-differences one was evaluated as $\delta F_{ERROR} = |\delta F_{ADJ} - \delta F_{FDM}|/|\delta F_{FDM}|$. Very low gradient error values were achieved, as seen in Fig.~\ref{fig2:fp_eta_results}(a), with an average value of $5.8\times10^{-6}$, comparable to those typically obtained in other successful adjoint optimizations \cite{Michaels.Andrew.2018_continuous_design}.

Two parameters that are potentially critical for the final optimization results are the quantum emitter position, and the starting geometry aspect ratio, defined as the length ($l$) over the width ($w$) of the design area $\Omega$, shown in Fig.~\ref{fig1:device}(a). We have analyzed the impact of these parameters on the values of $F_{p}$ and $\eta$ by optimizing the device for y-oriented electric dipole sources located at x-positions ranging from -0.5~$\mum$  to 0.5~$\mum$ (origin of the coordinate system at the center), and aspect ratios varying from 1.1 to 9.7. The design domain area was kept constant at 3.72~$\mum^{2}$, a value that fits into our simulation capabilities, whereas the domain length was increased from 2~$\mum$  to 6~$\mum$ as the aspect ratio was varied. 

Figure~\ref{fig2:fp_eta_results}(b) shows final optimization results obtained for various initial geometries and dipole locations. It is apparent that higher aspect ratios improved the final values of both $F_{p}$ and $\eta$. This might be attributed to the suppression of coupling into the various higher-order optical modes supported by the wider structures (see Supplementary Figure S1 and the discussion therein). The dipole position has a major role over $F_{p}$ and only slightly impacts the values of $\eta$ within this range. Except for the structure with the lowest aspect ratio, all of the optimized devices showed $F_{p}$ and $\eta$ values above 4 and 0.7, respectively. The highest values of $F_{p}$ and $\eta$ were 21 and 0.94, respectively, obtained for the highest aspect ratio structure (6 $\mum$ $\times$ 0.62 $\mum$) when the dipole source was located at position $\boldsymbol{r'} = (0.5, 0, 0)\:\mum$. Even better results can be obtained if the $\Delta_{QE}$ constraint is relaxed (Supplementary Table S2 and Figure S2).

Figure~\ref{fig2:fp_eta_results}(c) shows the evolution of $F_p$ and $\eta$ during the optimization process for the starting geometry with aspect ratio 9.7 and dipole located at $x=0.5\:\mum$. The overall coupling efficiency $\eta$ reaches a $0.80$ value after only a few iterations, even for moderate (< 5) $F_{p}$ values, as seen in Fig.~\ref{fig2:fp_eta_results}(c), and a steady-state value of $\eta = 0.94$ is reached after 200 iterations. The Purcell factor, however, shows an almost continuous enhancement until approximately iteration 200, at which point it makes a big leap, resulting in a final value of $21$ after 300 iterations. The leap may be due to the elimination of features that cause cavity scattering losses, and therefore lower quality factors, which may happen abruptly due to the permittivity binarization scheme discussed at the end of Section~\ref{section:methodology}. The small periodic fluctuations observed in $F_{p}$ and $\eta$ are attributed to the projection parameter increments that happen every 5 iterations. 

The evolution of the permittivity distribution $\boldsymbol{\varepsilon}(\rho, \boldsymbol{r})$ during the optimization process can be observed in Fig.~\ref{fig2:fp_eta_results}(d). It is apparent that the region within the circle of radius $\Delta_{QE}$ remains unchanged throughout, as required. Interestingly, after the first iteration, the spatial index distribution outside of the constraint circle presents a periodicity along x that is reminiscent of modal beating in the multimode waveguide. Indeed, the periodicity is consistent with the beat length $L_{2\pi}=\lambda/(n_1-n_3)\approx1\:\mum$ between the two TE modes in Fig.~\ref{fig1:device}(a), with effective indices $n_1=3.03$ and $n_3=2.10$, which are selectively excited by the dipole. Other guided modes supported by the multimode waveguide have no y-component of the electric field at the dipole location, and are therefore not excited. The spatial periodicity of the permittivity observed in the initial step can be observed throughout the rest of the optimization. The procedure converges to an index distribution that supports a cavity optical mode with electric field shown in the bottom panel of Fig.~\ref{fig2:fp_eta_results}(d), which features an apparent reduced modal volume and maximum intensity at the dipole position, and which funnels light into the fundamental optical mode supported by the output waveguide. The real part of the electric field components $E_{x}$ and $E_{y}$ present, respectively, a node and an anti-node at the dipole position, so a maximum Purcell factor is achieved for quantum emitters with y-oriented dipole moments.

Wavelength-dependent Purcell enhancement and coupling efficiency curves for the optimized geometry, shown in Fig. \ref{fig2:fp_eta_results}(e), were obtained by finite-difference time-domain (FDTD) simulations. The small discrepancies between the peak values of $F_{p}$ and $\eta$ with respect to the ones obtained in the optimization can be attributed mainly to the differences in the way the structure was discretized on the FDFD and FDTD electromagnetic solvers. A well-defined resonance centered at 936~nm can be seen in the $F_p$ curve, with a full-width at half-maximum (FWHM) of 2.6 nm, corresponding to a quality factor of 360. Importantly, the relatively low quality factor here compared to, for example, photonic crystal defect cavities is due to an efficient power extraction into the fundamental TE output waveguide mode. Indeed, the output waveguide extraction efficiency peaks at just below 0.9, and remains above 0.5 within a wavelength range of $>20$~nm. Comparable values of radiative rate enhancement and collection efficiency into a desired optical collection channel can be achieved with single-photon sources based on bullseye nanocavities~\cite{ref:Davanco_BE} for free-space, as opposed to guided-wave photon extraction. We emphasize also that broad resonance linewidths may be desirable to relax spectral alignment requirements between cavity and emitter, and to allow extraction of two or more spectrally separate transitions of the same emitter~\cite{Liu2019}.

\begin{figure*}[ht]
	\centering
	\includegraphics[scale=1.1]{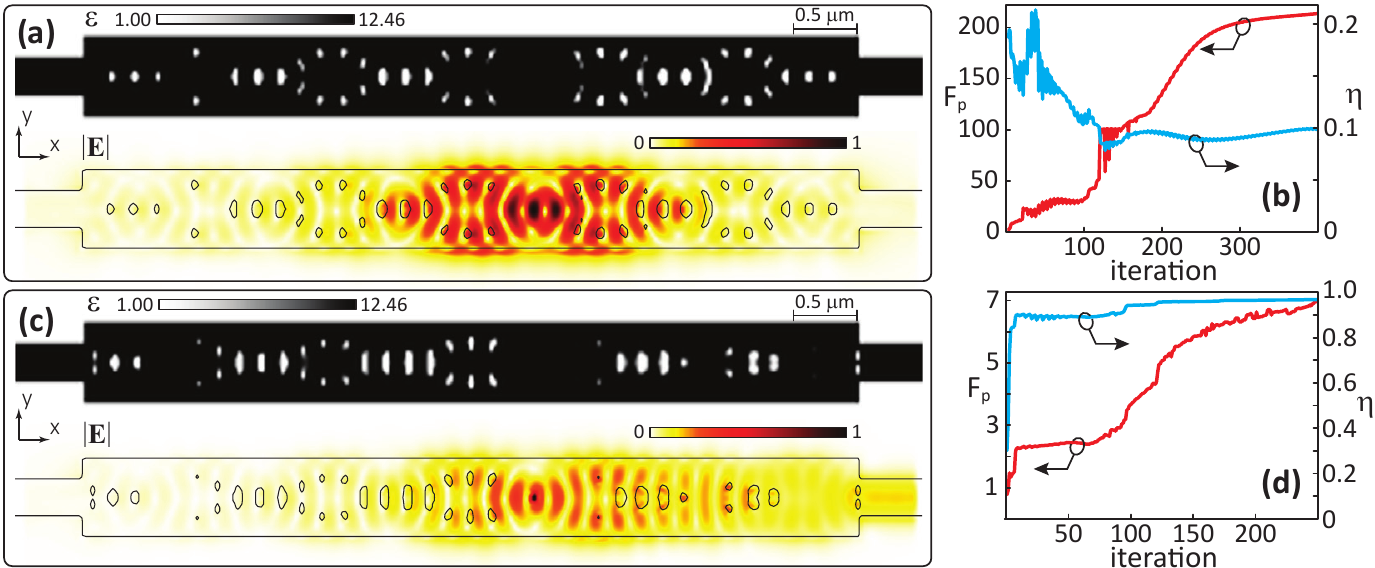}
	\caption{Optimization results for single-objective functions. Dielectric distribution ($\boldsymbol{\varepsilon}$) and absolute value of the electric field at the final iteration ($|E|$) normalized by its maximum value when the cost function was composed by only the objective $f_{1}$ \textbf{(a)} or $f_{2}$ \textbf{(c)}, as well as their respective values of $F_{p}$ and $\eta$ at each iteration in \textbf{(b)} and \textbf{(d)}. These results were obtained for a device with 9.7 aspect ratio (6 $\mum$ $\times$ 0.62 $\mum$) and the dipole source placed at the position $\boldsymbol{r'} = (0.5, 0, 0)\:\mum$.}
	\label{fig3:single_obj_results}
\end{figure*}

We also investigate the evolution of the optimization considering the same 6~$\mum$ $\times$ 0.62~$\mum$ and $\boldsymbol{r'} = (0.5, 0, 0)\:\mum$ device, when the cost function defined in Eq.~\ref{eq:opt_problem} includes only one objective at a time. A high Purcell factor nanocavity ($F_{p}$ = 220) was obtained when $f_{1}$ was minimized alone, as can be seen in Figs. \ref{fig3:single_obj_results}(a-b). Interestingly, the final index distribution clearly displays, to the left and right of the dipole, a similar periodic pattern as seen in Fig.~\ref{fig2:fp_eta_results}(d), which derives from modal beating. Such a pattern must act as an effective multimode mirror on both sides of the dipole, which enables the formation of the well confined cavity mode seen in the bottom panel of Fig.~\ref{fig3:single_obj_results}(a). The coupling efficiency into the output waveguide was very poor in this case ($\eta<0.1$), which is expected because output coupling reduces the quality factor of the cavity.

In the case where only the coupling efficiency FOM was considered, $\eta$ achieves a value of 0.96---higher than when considering both FOMs together---as shown in Fig.~\ref{fig3:single_obj_results}(d). $F_p$ also displays a considerable increase even though it is not explicitly optimized. As already demonstrated in previous works \cite{Yesilyurt.O.2021_sps_coupler}, an improved $\eta$ may indeed be obtained from an increase in $\beta$-factor that originates from a higher $F_{p}$. Nonetheless, the $F_{p}$ value obtained here is only $\approx2/3$ of that returned by the multi-objective optimization approach. We note also that the index distribution shown in Fig. \ref{fig3:single_obj_results}(c) to the left of the dipole is similar to that of the high $F_p$ cavity of Fig.~\ref{fig3:single_obj_results}(a). As discussed, this pattern provides high reflectivity to the multimode field generated by the dipole. To the right of the emitter, the pattern is significantly different and appears to lead simultaneously to partial reflectivity of the two dipole-excited modes and to matching of the multimode field into the output waveguide mode.

So far we have considered a high refractive index medium ($n>3.0$), which is very favorable for achieving high dipole coupling to confined optical modes~\cite{Friedler2009, ref:Davanco2}. A number of relevant quantum emitter material platforms do not feature such high refractive indices, however, and generally require more careful consideration. In Section S3 of the supplementary information, we show that our inverse-design methodology can also be successfully applied to one such material platforms, specifically Si$_{3}$N$_{4}$  ($n\approx2$), which may be employed to create single-photon sources based on hexagonal boron nitride (hBN) color centers~\cite{Yesilyurt.O.2021_sps_coupler}. 

The optimized geometries presented above feature complex shapes of very small dimensions which pose challenges for fabrication. Adding extra shape optimization steps with a modified cost function to efficiently enforce realistic fabrication constraints is a potential path to circumventing this issue~\cite{vercruysse_analytical_2019,Piggott.A.Y.2020_inv_des_foundries}. Alternatively, more sophisticated density-based projection filters to control the feature sizes~\cite{sigmund_manufacturing_2009,hammond_photonic_2021} may be applied alongside our methodology. While such possibilities go beyond the scope of the present work, they will be explored in future work.

Our results highlight the fact that inverse design is capable of generating sources with $\eta$ and $F_p$ values that are competitive with those of sources designed based on pre-established waveguide or cavity geometries~\cite{katsumi_transfer-printed_2018}. The latter approach starts with optimization of a pre-determined waveguide~\cite{mouradian_scalable_2015,davanco_heterogeneous_2017,daveau_efficient_2017} or cavity~\cite{katsumi_quantum-dot_2019,wang_micropillar_2020}, targeting $\beta$, $F_p$, or quality factor. Optimization of the access channel coupling efficiency $\eta$ then follows, in which a trade-off between $F_p$ and $\eta$ must be reached. In this process, minimization of parasitic optical modes must be minimized, which may be challenging depending on the starting geometry~\cite{katsumi_transfer-printed_2018}. Our results indicate that inverse design addresses all such considerations simultaneously, likely creating more possibilities for favorable trade-offs to be achieved. Remarkably, the optimization furthermore obeys an imposed geometric constraint---keeping etched sidewalls far from the emitter---that significantly complicates the design of small mode-volume cavities, and would likely lead to difficult trade-offs when using conventional optimization procedures.

\section{Conclusion}

We have demonstrated a multi-objective inverse-design methodology that can be applied to the design and optimization of quantum emitter based single-photon sources, aiming to achieve simultaneously high overall quantum emitter coupling efficiencies ($\eta$) and Purcell factors ($F_{p}$). Our results indicate that the technique can produce comparable or better performance than conventional design approaches, provided both objectives are targeted simultaneously. In particular, a high $F_p$ is maintained even with simultaneously high $\eta$, a non-trivial result given the interdependence of the two parameters. To produce these results, we relied on an adjoint source formulation that takes into account the geometry-dependent variation of the total dipole-emitted power. Such a consideration, seldom made in prior work, is shown to be critical for reaching optimal trade-offs in quantum emitter light extraction optimization. 

We have illustrated our method with the design of a compact ($3.72$ $\mum^{2}$) on-chip single-photon source based on InAs quantum dot emitters in a GaAs host. The design procedure yielded high (0.94 peak) and broadband (FWHM$\approx20$~nm) single-photon coupling efficiency into an on-chip waveguide mode, and a Purcell factor of as much as 21, for a quantum dot in a geometry that featured no etched sidewalls at distances closer than 300 nm from its position. Such results are competitive with those obtained through a conventional design approaches starting from pre-established waveguide or cavity geometries, and are achieved in spite of an imposed geometrical constraint. Indeed, such a combination of desirable characteristics is unprecedented for this particular class of quantum emitters, and are likely also achievable in other materials systems. 

Our results furthermore illustrate how inverse-design procedures may lead to geometries whose workings may yet be grasped using well-known electromagnetic concepts, and may even provide reasonable intuition for novel designs. We anticipate that our contribution will enable the inverse design of a wide class of quantum photonic devices with functionality enabled by single solid-state quantum emitters, such as quantum logic gates and spin-photon interfaces, for which efficient optical access channels to the emitter are a necessity.  

\noindent \textbf{Acknowledgements:}
We thank the S\~{a}o Paulo Research Foundation (FAPESP) Grant 2021/10249-2 for the financial support and the Brazilian National Laboratory of Scientific Computation (LNCC) for the high performance computing (HPC) resources.

\noindent \textbf{Data availability:}
Data and simulation files are openly available in the Zenodo repository at http://doi.org/10.5281/zenodo.6463675 \cite{E.G.Melo.2022_zenodo}.


\setcounter{figure}{0}
\setcounter{equation}{0}
\setcounter{section}{0}

\renewcommand{\thesection}{S\arabic{section}}
\renewcommand{\thetable}{S\arabic{table}}
\renewcommand{\thefigure}{S\arabic{figure}}

\newpage

\begin{center} {{\bf \large Supplementary Information: Multi-Objective Inverse Design of Solid-state Quantum Emitter Single-photon Sources}}\end{center}



\section{Impact of design domain aspect ratio}
\begin{figure*}[!h]
	\centering
	\includegraphics[scale=1.00]{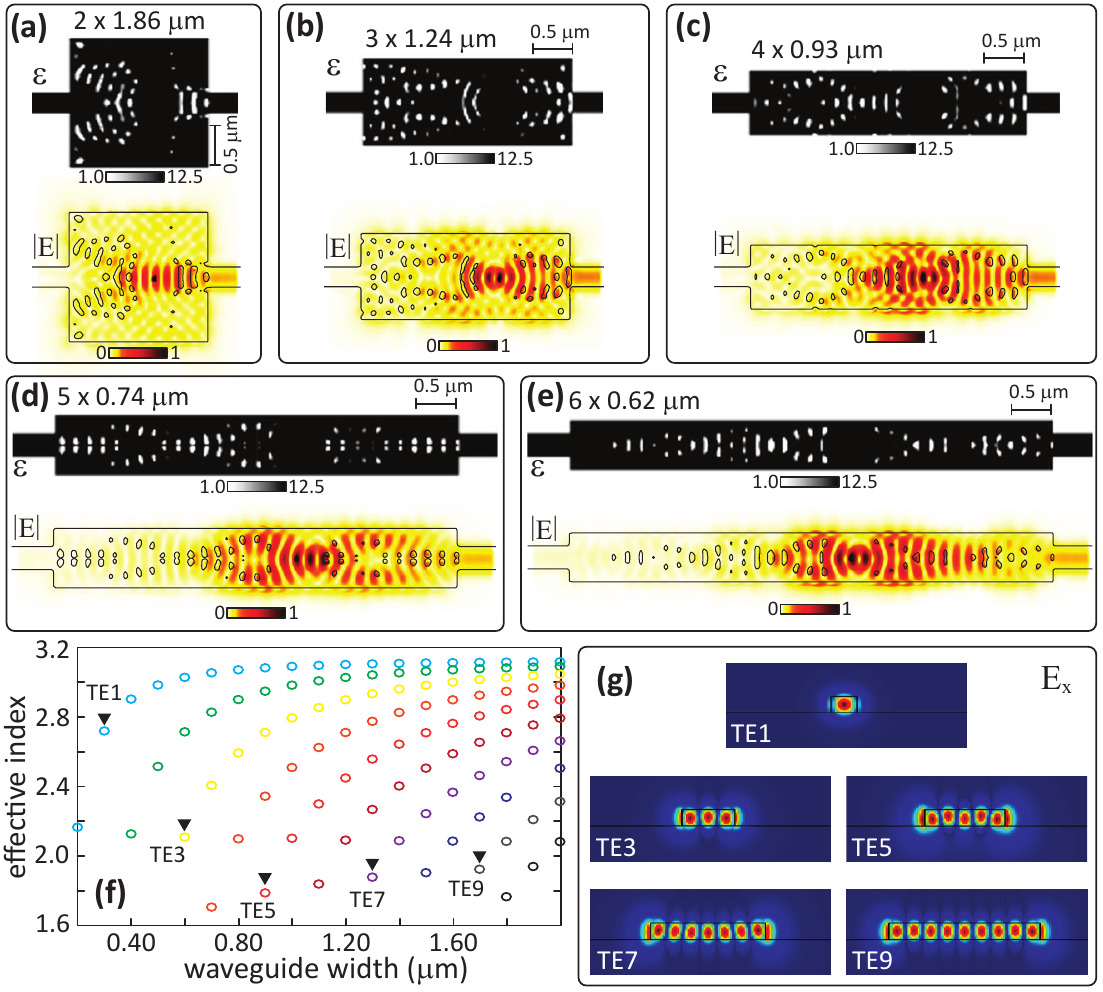}
	\caption{Optimization results for geometries with different aspect ratios. Dielectric distribution ($\boldsymbol{\varepsilon}$) and absolute value of the normalized electric field ($|E|$) at the final iteration of the runs that resulted in optimized values of $F_{p}$ and $\eta$ for devices with aspect ratios of 1.1 \textbf{(a)},  2.4 \textbf{(b)}, 4.3 \textbf{(c)}, 6.8 \textbf{(d)}, and 9.7 \textbf{(e)}; \textbf{(f)} Mode effective index values with respect to the waveguide width for TE polarization; \textbf{(g)} x-component of the electric field distribution ($E_{x}$) for the optical modes indicated by the black triangles shown in (f).}
	\label{fig:S1}
\end{figure*}

Dielectric distributions ($\boldsymbol{\varepsilon}$) and corresponding electric field amplitudes ($|E|$) with simultaneously optimized $F_{p}$ and $\eta$ are shown in Fig.~\ref{fig:S1}(a-e) for devices with aspect ratios (length over width) of 1.1, 2.4, 4.3, 6.8, and 9.7, respectively. The optimized values of $F_{p}$ and $\eta$ are summarized in Table \ref{tab:tab1}.

As discussed in the main text, the optimizations have returned higher values of Purcell factor ($F_{p}$) and overall coupling efficiency ($\eta$) for increasing aspect ratios. This tendency could be potentially related with the difficulties created for the optimization algorithm when the dipole emission can couple into many higher order waveguide modes, thereby giving rise to complex modal interference patterns. Observing Fig.~\ref{fig:S1}(f) we can see that the structures with 0.62 $\mum$ and 0.75 $\mum$ width, which have the larger values of $F_{p}$ and $\eta$, support only 2 transverse electric (TE) optical modes with even symmetry, shown in Fig.~\ref{fig:S1}(g). Three even-symmetric TE modes can be excited in the devices with 0.93~$\mum$ and 1.24 $\mum$ of width, and 5 modes are supported by the largest device. Potentially, decreasing the aspect ratio while also increasing the design region area could generate better results, however, this analysis was not possible within our computational capabilities.
\begin{table}[!h]
\centering
\begin{tabular}{|c|c|c|c|}
	\hline
	aspect ratio  & dimensions & $F_{p}$ & $\eta$ \\
	\hline\hline
	1.1 & 2 $\mum$ x 1.86 $\mum$ & 6 & 0.75\\
	\hline
	2.4 & 3 $\mum$ x 1.24 $\mum$ & 11 & 0.89\\
	\hline
	4.3 & 4 $\mum$ x 0.93 $\mum$ & 14 & 0.88\\
	\hline
	6.8 & 5 $\mum$ x 0.74 $\mum$ & 18 & 0.93\\
	\hline
	9.7 & 6 $\mum$ x 0.62 $\mum$ & 21 & 0.94\\
	\hline
\end{tabular}
\caption{Different values of $F_{p}$ and $\eta$ obtained for devices with a design region possessing different aspect ratios.}
\label{tab:tab1}
\end{table}
\newpage
\section{Impact of constraint region diameter }
\begin{figure*}[ht]
	\centering
	\includegraphics[scale=1.00]{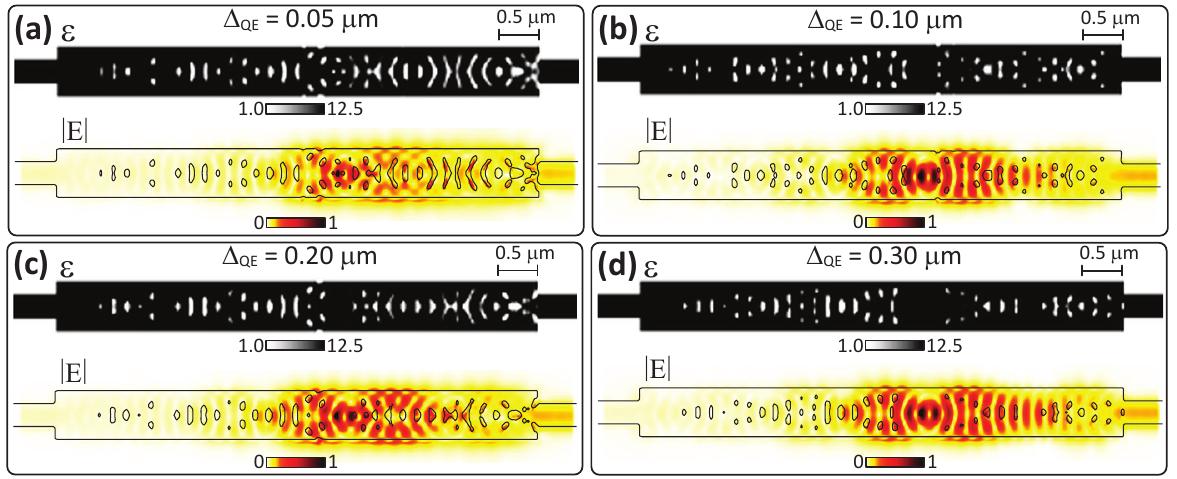}
	\caption{Optimization results for varying $\Delta_{QE}$. Dielectric distribution ($\boldsymbol{\varepsilon}$) and absolute value of the electric field ($|E|$) normalized by its highest value at the final iteration for optimizations performed in devices with 6 $\mum$ $\times$ 0.62 $\mum$ design regions and radius $\Delta_{QE}$ with values of 0.05 $\mum$ \textbf{(a)},  0.10 $\mum$ \textbf{(b)}, 0.20 $\mum$ \textbf{(c)}, and 0.30 $\mum$ \textbf{(d)}.}
	\label{fig:S2}
\end{figure*}

The absolute value of the electric field ($|E|$) and the dielectric distributions ($\boldsymbol{\varepsilon}$) at the final iteration of the optimizations performed with varying values of $\Delta_{QE}$, are shown in Fig.~\ref{fig:S2}(a-d), respectively for $\Delta_{QE} = (0.05,\:0.10,\:0.20$, and $0.30)\:\mum$. $\Delta_{QE}$ is the radius of a circular area around the quantum emitter position where perturbations to the dielectric distribution are forbidden (see Fig.~1 of the main text). All these optimizations were performed considering the device with a 6$\mum$ $\times$ 0.62 $\mum$ design region.
As summarized in Table \ref{tab:tab2}, no clear trends can be observed for $F_{p}$ and $\eta$ with respect to $\Delta_{QE}$, however, it is apparent that better results could be obtained for specific conditions. For instance, for $\Delta_{QE} = 0.10\:\mum$, $F_{p} = 24.2$ and $\eta = 0.95$ were achieved, which are slightly higher values than those for $\Delta_{QE} = 0.30\:\mum$ described in the main text.
\begin{table}[!h]
\centering
\begin{tabular}{|c|c|c|c|}
	\hline
	$\Delta_{QE}$  & $F_{p}$ & $\eta$ \\
	\hline\hline
	0.05 $\mum$ & 13.3 & 0.84\\
	\hline
	0.10 $\mum$ & 24.2 & 0.95\\
	\hline
	0.20 $\mum$ & 17.6 & 0.91\\
	\hline
	0.30 $\mum$ & 21 & 0.94\\
	\hline
\end{tabular}
\caption{Different values of $F_{p}$ and $\eta$ obtained for devices optimized with varying radius $\Delta_{QE}$.}
\label{tab:tab2}
\end{table}
\newpage
\section{Optimization of $\SiN$ geometry}
\begin{figure*}[ht]
	\centering
	\includegraphics[scale=1.00]{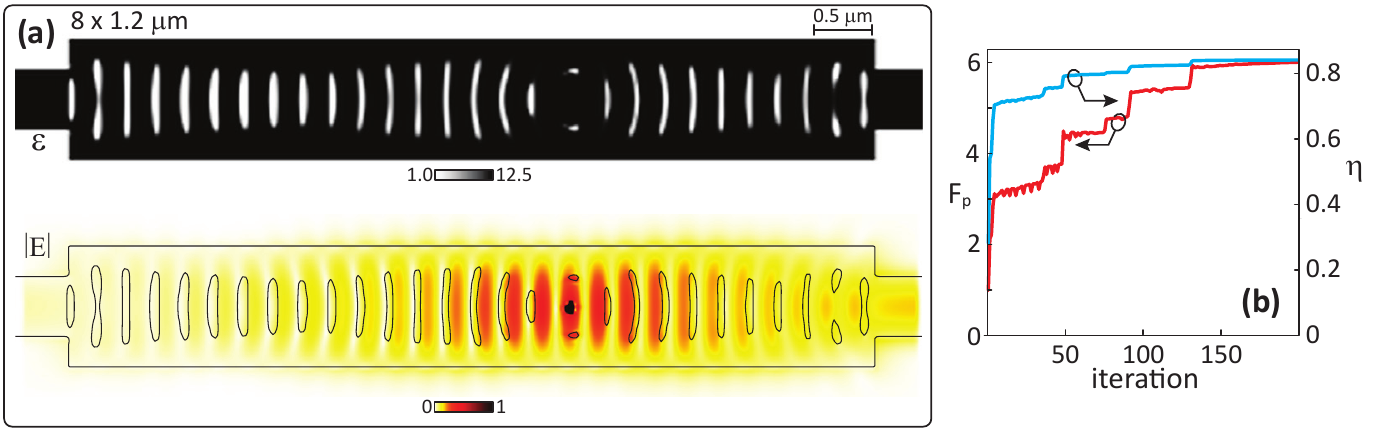}
	\caption{Optimization result for a mid-range refractive index ($n\approx2$) material platform. \textbf{(a)} Dielectric distribution ($\boldsymbol{\varepsilon}$) and absolute value of electric field ($|E|$) normalized by its highest value at the final iteration of an optimization performed considering a device composed of a 250 nm thick Si$_{3}$N$_{4}$ thin film over a SiO$_{2}$ substrate; \textbf{(b)} Values of $F_{p}$ and $\eta$ at each iteration.}
	\label{fig:S3}
\end{figure*}

Here we present the application of the proposed methodology for the inverse design of nanophotonic solid-state quantum emitter single-photon sources in a mid-range ($n\approx2$) refractive index material platform, composed of a 250 nm thick Si$_{3}$N$_{4}$ thin film over a SiO$_{2}$ substrate. In this case, the design region is 8 $\mum$ in length and 1.20 $\mum$ wide, and the output waveguide has a 0.25 $\mum$ $\times$ 0.60 $\mum$ cross section, which ensures a single transverse-electric mode for the extraction of the generated single photons. As in the case studied in the main text, the quantum emitter was also modeled as a y-oriented electric dipole source embedded in the center of the Si$_{3}$N$_{4}$ thin film, but, in this case, displaced by 1.0 $\mum$ along the x-direction. The dielectric distribution ($\boldsymbol{\varepsilon}$) at the final iteration process and the absolute value of the electric field ($|E|$) are shown in Fig.~\ref{fig:S3}(a). The geometry resembles a nanobeam cavity with elliptical holes and a well defined periodicity. After 200 iterations,  $F_{p} = 6$ and $\eta = 0.83$ were obtained, as shown in Fig.~\ref{fig:S3}. Even though these values are lower than the ones presented on the main text for a high-index material platform, the Purcell factor and overall coupling efficiency are higher than those reported for a similar material platform in ref.~\cite{Yesilyurt.O.2021_sps_coupler}, where the objective function included only the coupling efficiency figure-of-merit.

\end{document}